\documentclass[runningheads]{llncs}
\usepackage{float}
\usepackage{bbding}
\usepackage{booktabs}    
\usepackage{graphicx}

\usepackage[misc]{ifsym}
\usepackage[namelimits]{amsmath} 
\usepackage{amssymb}             
\usepackage{amsfonts}            
\usepackage{mathrsfs}            
\usepackage{multirow}
\begin{document}
\title{CANet: Channel Extending and Axial Attention Catching Network for Multi-structure Kidney Segmentation}
\titlerunning{CANet.}
%
%
\author{Zhen-Yu Bu\orcidID{0000-0001-8360-0535} \and
Kai-Ni Wang\orcidID{0000-0002-4000-188X} \and
Guang-Quan Zhou$^{(\textrm{\Letter})}$\orcidID{0000-0002-6467-3592}}

\authorrunning{Z. Bu et al.}
%
\institute{Monash University, Southeast University, School of Biological Science and Medical Engineering, Nanjing, China\\
\email{guangquan.zhou@seu.edu.cn}\\}
\maketitle              
\begin{abstract}
Renal cancer is one of the most prevalent cancers worldwide. Clinical signs of kidney cancer include hematuria and low back discomfort, which are quite distressing to the patient. Some surgery-based renal cancer treatments like laparoscopic partial nephrectomy relys on the 3D kidney parsing on computed tomography angiography (CTA) images. Many automatic segmentation techniques have been put forward to make multi-structure segmentation of the kidneys more accurate. The 3D visual model of kidney anatomy will help clinicians plan operations accurately before surgery. However, due to the diversity of the internal structure of the kidney and the low grey level of the edge. It is still challenging to separate the different parts of the kidney in a clear and accurate way. In this paper, we propose a channel extending and axial attention catching Network(CANet) for multi-structure kidney segmentation. Our solution is founded based on the thriving nn-UNet architecture. Firstly, by extending the channel size, we propose a larger network, which can provide a broader perspective, facilitating the extraction of complex structural information. Secondly, we include an axial attention catching(AAC) module in the decoder, which can obtain detailed information for refining the edges. We evaluate our CANet on the KiPA2022 dataset, achieving the dice scores of 95.8\%, 89.1\%, 87.5\% and 84.9\% for kidney, tumor, artery and vein, respectively, which helps us get fourth place in the challenge.
\end{abstract}
\section{Introduction}
Renal cancer is one of the most prevalent cancers in the world, ranking 13th globally and having caused more than 330,000 new cases worldwide. \cite{chow2010epidemiology} Laparoscopic partial nephrectomy, a dependable and successful treatment for kidney cancer, has gained widespread use.\cite{ljungberg2019european} However, laparoscopic partial nephrectomy requires a well-defined segmented anatomy of the kidney. \cite{he2021meta,shao2011laparoscopic,shao2012precise} Since annotating kidney structures manually initially needs trained professionals, which is both time-consuming and costly. It is essential to develop a technique for automatic multi-structure kidney segmentation.

Nowadays, numerous deep learning techniques have been developed for medical image segmentation, localization and detection.\cite{zhou2021learn,wang2022awsnet,ronneberger2015u,cheng2006nanotechnologies} Since the diversity in the internal structure of the kidney (see in Figure 1), it it difficult to clearly segment the structure of the kidney to a certain extent. Meanwhile, the low gray level of the edge is also a problem.\cite{he2021meta} There are numerous techniques for segmenting the kidney's structure.\cite{wang2020tensor,jin20163d,he2019dpa,lin2006computer} However, these methods only target a single region, such as the kidney, and do not segment the four components concurrently. This cannot supply sufficient information for the downstream tasks. There are progressive multi-structured divisional approaches, however their effects vary and they are not favorable.\cite{li2018segmentation,taha2018kid}

In this paper, we propose a simple and effective channel extending and axial attention catching Network(CANet) based on the nn-UNet backbone for multi-structure kidney segmentation. Straightforwardly, we enlarge the size of the filters along with adding axial attention module in the decoder phase. Meanwhile, some basic setting of nn-UNet has been modified to get the final results. We use kidney multi-structure data (KiPA22 dataset \cite{he2021meta,shao2011laparoscopic,shao2012precise,he2020dense}) to evaluate the proposed method. We demonstrate experimentally that despite the fact that our model invariably increases training time, it generates substantial benefits in our experiment.

\begin{figure}[H]
\begin{center}
\vspace{-1em}  
\includegraphics[width=1\textwidth]{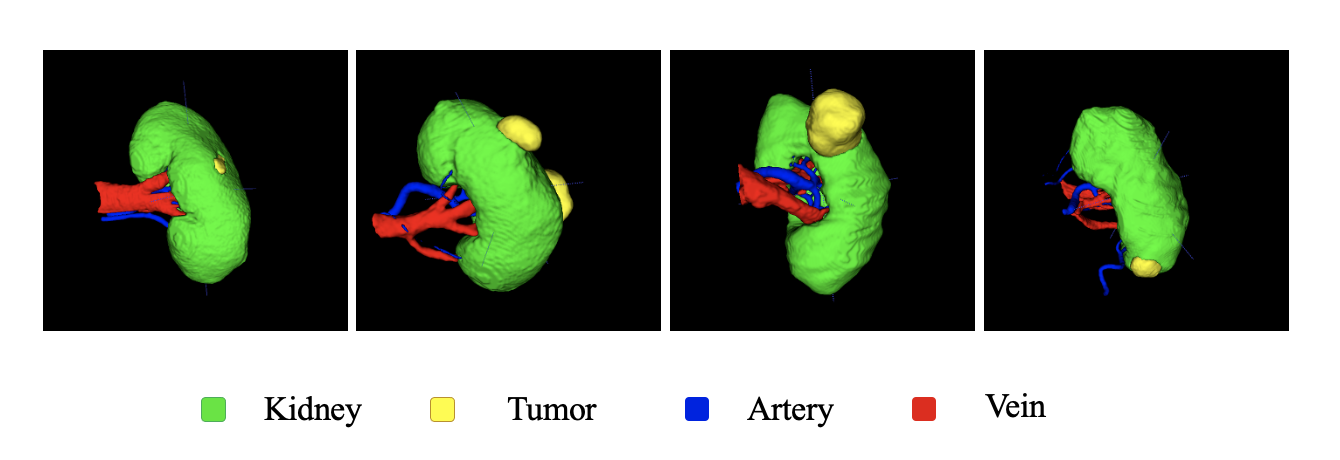}
\label{network.eps}
\caption{The shape of the kidneys, tumors, and the fact that the blood vessels aren't very thick are just some of the problems shown in the figure above. The yellow ones are tumors, the green ones are kidneys, and the red and blue ones are veins and arteries.}
\vspace{-1em}  
\end{center}
\end{figure}

\section{Methods} 
Figure 2 illustrates the pipeline of our proposed CANet. First, we preprocess the data, then we transform the input 3D data into the corresponding form. Then we feed the processed data into our CANet, and finally we obtain the final result using post-processing technology.
\begin{figure}[H]
\begin{center}
\vspace{-1em}  
\includegraphics[width=1\textwidth]{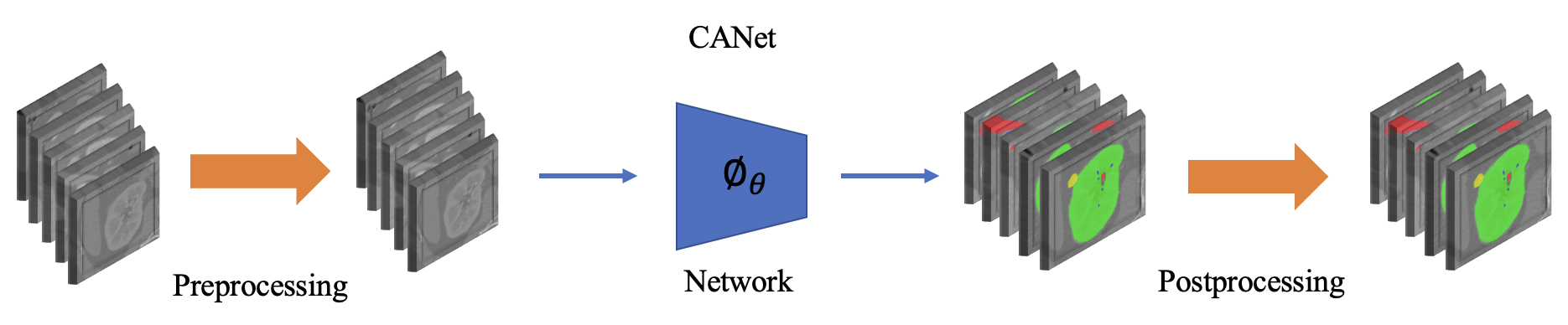}
\label{network.eps}
\caption{An overall pipeline of our proposed CANet. This pipeline mainly includes two parts, pre/post processing and the core deep learning network.}
\vspace{-1em}  
\end{center}
\end{figure}

\subsection{Preprocessing}
The preprocessing method uses resample and normalization. There are data with different spacing, which is automatically normalized to the median spacing of all data spacing by default. The original data uses third-order spline interpolation and mask uses nearest neighbor interpolation. We count the HU value range of the pixels in the mask in the entire data set and then clip out the HU value range of the [0.05, 99.5] percentage range. Finally the z-score method for normalization is utilized. After these steps, the standard input of nn-UNet is able to be fed to the model for training.

\begin{equation}
z=\frac{x-\mu}{\sigma}
\end{equation}
where $\mu$ is the mean and $\sigma$ is the variance. It can convert data of different magnitudes into a unified Z score for comparison.

\begin{figure}[H]
\begin{center}
\vspace{-1em}  
\includegraphics[width=0.8\textwidth]{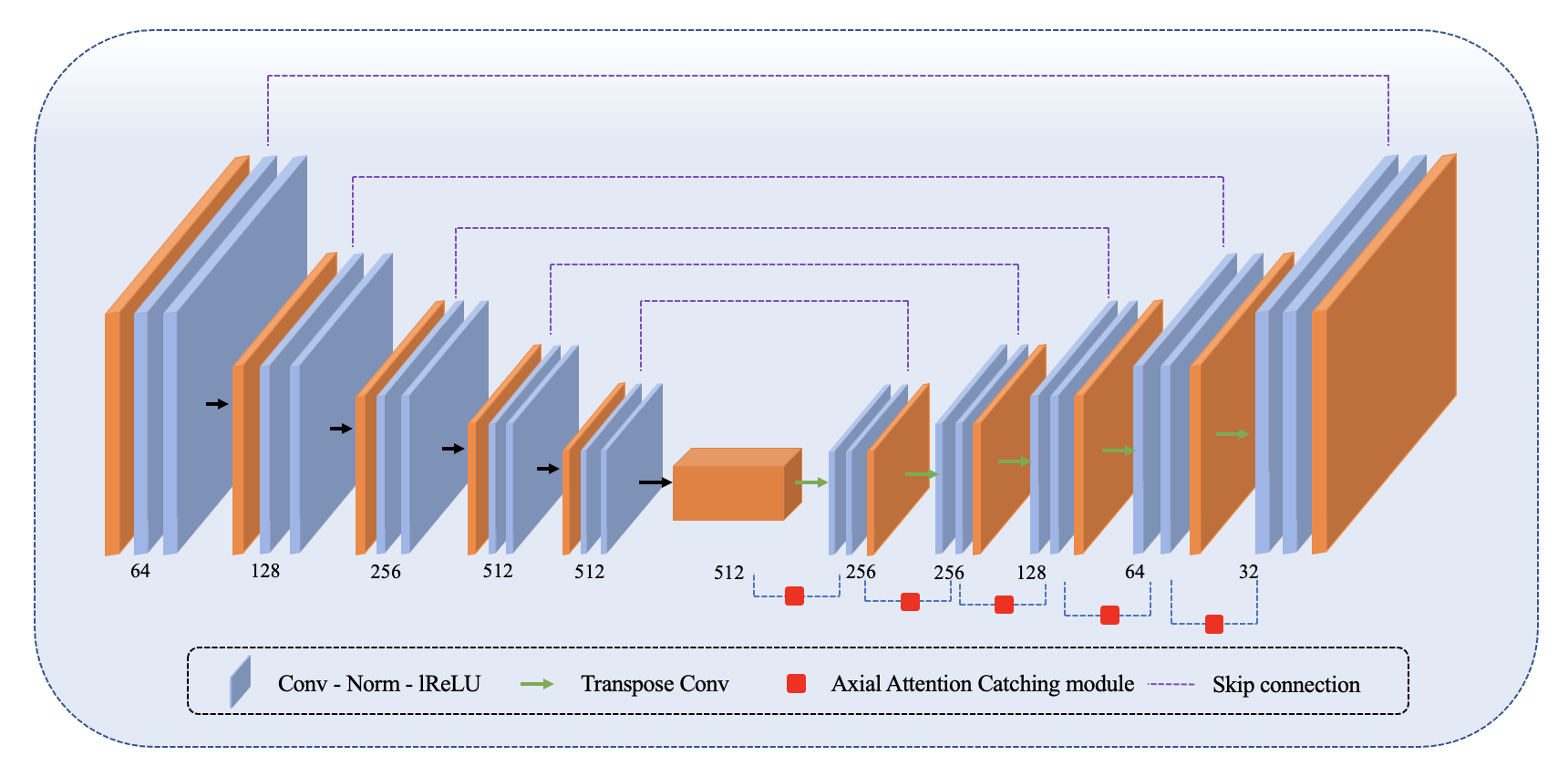}
\label{network.eps}
\caption{An overview of our proposed CANet, which is based on the 3D full resolution U-net for kidney multi-structure segmentation. }
\vspace{-1em}  
\end{center}
\end{figure}

\subsection{nn-UNet backbone}
The 3D full resolution nn-UNet is used as our backbone.\cite{isensee2018nnu} The network's structure is encoder-decoder with a skip connection. No new structures, such as residual block, dense connection, or attention mechanism, are added to the original UNet. Instead, it emphasized data preprocessing, training, inference and post-processing.

\subsection{Channel Extending (CE)} 
We doubled the number of filters in the encoder while leaving the number of filters in the decoder unchanged compared to the original nn-UNet structure(3D full resolution) has maximum number of filter is 320.\cite{isensee2018nnu} Taking into account the magnitude of the data, boosting the network's capacity will give a larger view, hence simplifying the extraction of intricate structural information. The maximum number of filter is increased to 512. This work was first used in BraTS21 for brain tumor segmentation by Huan Minh Luu.\cite{luu2022extending} We do not apply group normalization in comparison to their work, instead.  

\subsection{Axial Attention Catching Module (AAC)} 
We insert axial attention catching module (AAC module) in the decoder to enhance the representation of the edge feature in order to extract sharper edge information. Figure 4 demonstrates that following the transposed convolution, position embedding is performed on the input, followed by vertical, horizontal, and depth operations, and then the output is concatenated. AAC module is a technique that helps lower the computational complexity\cite{wang2020axial} along with capturing the global information. For more specific, the axial-attention method minimizes the computational complexity of this form by first performing self-attention in the vertical direction and then doing self-attention in the horizontal direction and depth direction. By using the AAC module, the ability to extract edge information is enhanced in this way.

\begin{figure}[H]
\begin{center}
\vspace{-1em}  
\includegraphics[width=0.8\textwidth]{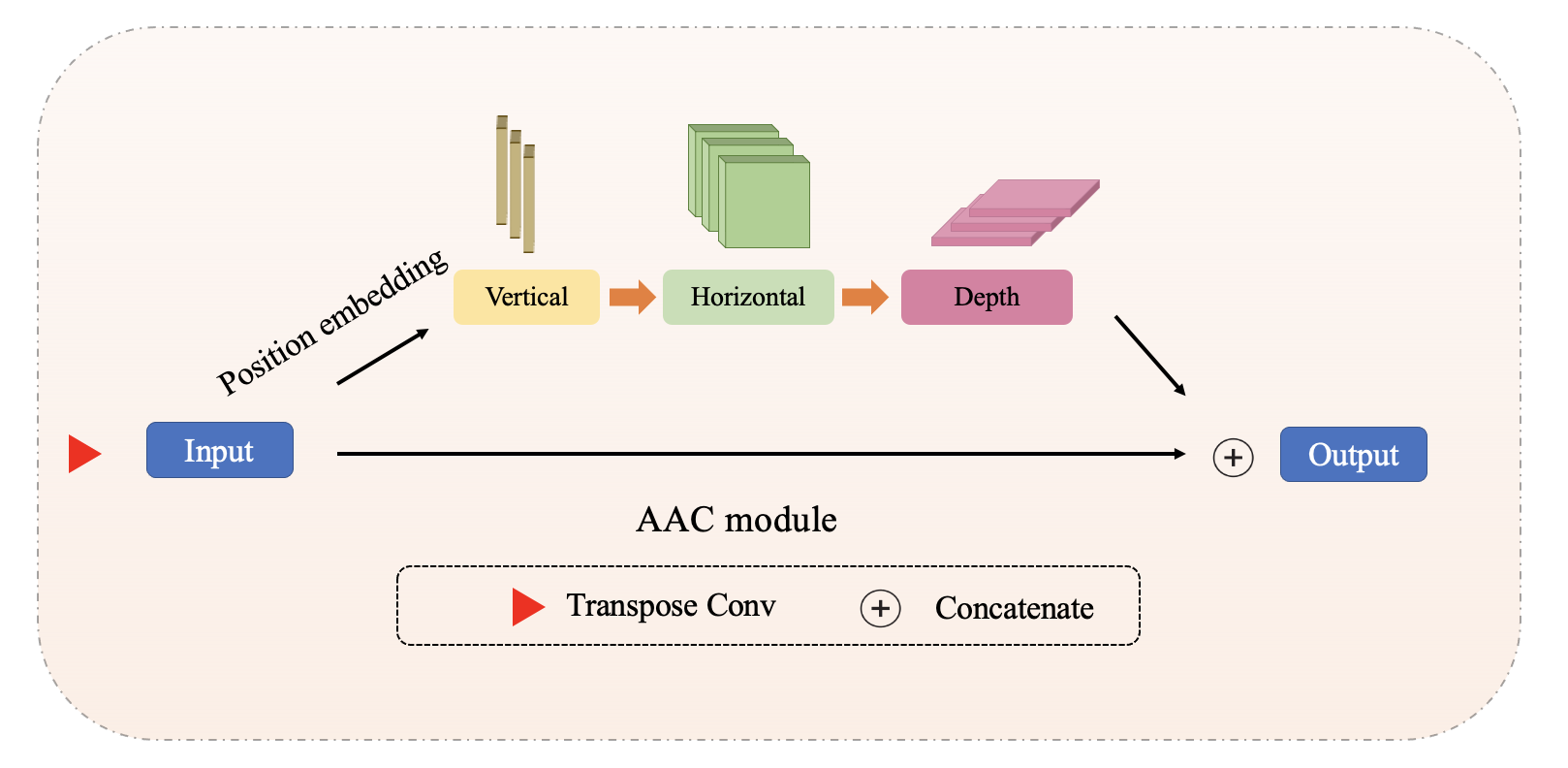}
\label{network.eps}
\caption{Visualization of axial attention catching(AAC) module. After transposed convolution, the input has gone through three self-attentions in vertical, horizontal and depth, and finally concatenate to get the final result.}
\vspace{-1em}  
\end{center}
\end{figure}

\subsection{Loss function}
In our CANet, we use a combination of dice loss and cross-entropy loss as our loss function.

\begin{equation}
    \mathcal{L}_{ {total }}=\mathcal{L}_{ {dice }}+\mathcal{L}_{C E}
\end{equation}\\
The first item is dice loss. We can calculate the dice coefficient first.
\begin{equation}
    D S C\left(y, y^{\prime}\right)=\frac{2 *\left|y \cap y^{\prime}\right|+\text { smooth }}{|y|+\left|y^{\prime}\right|+\text { smooth }}
\end{equation}\\
Then we can use the formula below to achieve the final Dice loss.
\begin{equation}
    \mathcal{L}_{dice} = 1 - DSC
\end{equation}\\
where $y$ indicates the ground truth and $y^{'}$ represents the predicted result.\\
Cross-entropy loss is represented by the following formula.
\begin{equation}
    \mathcal{L}_{CE}=-[y \log \hat{y}+(1-y) \log (1-\hat{y})]
\end{equation}\\
where $y$ and $y^{'}$ can be considered as ground truth and predict results respectively.


\section{Experiments}
\subsection{Data preparation}
In Kidney PArsing Challenge 2022, original images are CT scans from 130 patients (70 for training, 30 for opening test and 30 for closed test).\cite{he2021meta,shao2011laparoscopic,shao2012precise,he2020dense} All of these data was acquired on a Siemens dual-source 64-slice CT scanner and all of these data is in $nii.gz$ format. 
A sample corresponds to a ground truth. The kidneys range in size from 73.73ml to 263.02ml, whereas the tumors range from 2.06ml to 646.68ml. Meanwhile, there are also 30 samples for the open test phase and 30 samples for the closed test phase.

\subsection{Implementation details}
In the training stage, we divide the 70 training data into two parts, 56 for training and the following 14 for validation. Before training, data undergoes preprocessing, such as scaling, random rotation, elastic deformation, gamma scaling, etc. We set the batch size to 2. We employ 300 epochs for each training fold as opposed to 1000 epochs in the origin nn-UNet, which saves a great deal of time and yields excellent results. In our experiment, our training are on NVIDIA RTX 3090 GPU with 24GB VRAM, which takes about 3 minutes for one epoch. Finally, we also employ 5-fold cross-validation to improve performance. 
At the inference stage, we first do the preprocessing as training, and then fed the input into the model directly. After that, we also run the postprecessing method to link some unconnected areas.

\section{Results and discussions}
\subsection{Ablation experiments}
We do some ablation experiments to confirm the utility of channel extending mechanism and the AAC module in our CANet. In order to perform a fair experiment, all ablation experiments are conducted under same experiment setup. Here, CE refers to the channel extending mechanism and AAC means axial attention catching module. 
We test only CE and only AAC separately, and then combine them for testing. As shown in Table 1, the segmentation results were evaluated by dice score. The results show that both of our contributions are effective for improving the performance of the segmentation task.

\begin{table*}[]
\scriptsize
\centering
\caption{Ablation experiments examine the significance of our discoveries. The channel extension mechanism and AAC module have been shown to be beneficial in the segmentation of several kidney structures.}
\label{Tab03}
\begin{tabular}{cccccc}
\toprule
\multirow{1}*{CE} & \multirow{1}*{AAC} & \multicolumn{4}{c}{DSC(\%)}
\\
&   
&  Kidney      
&  Tumor  
&  Artery
&  Vein\\      
\midrule
\CheckmarkBold &    & 95.8 & 88.6 & 87.5 & 84.5\\
  &   \CheckmarkBold &  95.7  & 88.7 & 87.3 & 84.8\\
\CheckmarkBold & \CheckmarkBold & 95.8 & 89.1 & 87.5 & 84.9\\
\bottomrule
\end{tabular}
\end{table*}

Table 1 presents the quantitative results for the multi-structure kidney segmentation results on four parts, including kidney, tumor, artery and vein. It demonstrates that the channel extending mechanism can get 0.4\%  and AAC module are effective. Compared to the nn-UNet backbone, the channel extending mechanism can get 0.4\%, 0.7\%, 0.4\% dice score improvement on kidney, artery and vein respectively. Meanwhile, the AAC module can outperform 0.3\%, 0.6\%, 0.7\% dice score on kidney, artery and vein respectively. When combining both of them, the final CANet can get 0.4\%, 0.2\%, 0.8\%, 0.8\% dice score on kidney, tumor, artery and vein respectively. These data demonstrate the effectiveness of our contributions.

\subsection{Comparison experiments}
We also compare our method with several classical deep learning methods like DenseBiasNet, MNet, 3D Unet. Table 2 demonstrates the quantitive results of our comparison experiments. The number displayed in bold signifies the highest performance in each respective column. In almost every area, our model has the best performance.

\begin{table*}[]

\scriptsize

\centering

\caption{The quantitative assessment reveals that our CANet is superior for our 3D multi-structure kidney segmentation task. On all four renal structures, our CANet outperforms the comparative approaches (DenseBiasNet\cite{he2019dpa}, MNet\cite{dong2022mnet}, 3D U-Net\cite{cciccek20163d}, and nn-UNet\cite{isensee2018nnu}).}

\label{Tab03}

\begin{tabular}{ccccccc}

\toprule

\multirow{1}*{Methods} & \multicolumn{3}{c}{Kidney} & \multicolumn{3}{c}{Tumor} \\

\cmidrule(r){2-4} \cmidrule(r){5-7}

&  $DSC(\%)$      
&  $HD(mm)$   
&  $AVD(mm)$
&  $DSC(\%)$      
&  $HD(mm)$   
&  $AVD(mm)$\\

\midrule

DenseBiasNet   & 94.6 & 23.89 & 0.79 & 79.3 & 27.97 & 4.33 \\

MNet           & 90.6 & 44.03 & 2.16 & 65.1 & 61.05 & 10.23 \\
        
3D U-Net       & 91.7 & 18.44 & 0.75 & 66.6 & 24.02 & 4.45 \\

nn-UNet        & 95.4 & 18.56 & 0.78 & 88.9 & 14.68 & 1.50 \\

\textbf{CANet}          & \textbf{95.8} & \textbf{16.93} & \textbf{0.46} & \textbf{89.1} & \textbf{10.22} & \textbf{1.23} \\

\bottomrule

\end{tabular}

\end{table*}

\begin{table*}[]

\scriptsize

\centering


\label{Tab03}

\begin{tabular}{ccccccc}

\toprule

\multirow{1}*{Methods} & \multicolumn{3}{c}{Artery} & \multicolumn{3}{c}{Vein} \\

\cmidrule(r){2-4} \cmidrule(r){5-7}

&  $DSC(\%)$      
&  $HD(mm)$   
&  $AVD(mm)$
&  $DSC(\%)$      
&  $HD(mm)$   
&  $AVD(mm)$\\

\midrule

DenseBiasNet   & 84.5 & 26.67 & 1.31 & 76.1 & 34.60 & 2.08 \\

MNet           & 78.2 & 47.79 & 2.71 & 73.5 & 42.60 & 3.06 \\
        
3D U-Net       & 71.9 & 22.17 & 1.10 & 60.9 & 22.26 & 3.37 \\

nn-UNet        & 86.7 & 23.24 & 0.65 & 84.1 & 20.04 & 1.18 \\

\textbf{CANet}         & \textbf{87.5} & \textbf{15.12} & \textbf{0.34} & \textbf{84.9} & \textbf{11.87} & \textbf{0.67} \\

\bottomrule

\end{tabular}

\end{table*}

\begin{figure}[H]
\begin{center}
\vspace{-1em}  
\includegraphics[width=1\textwidth]{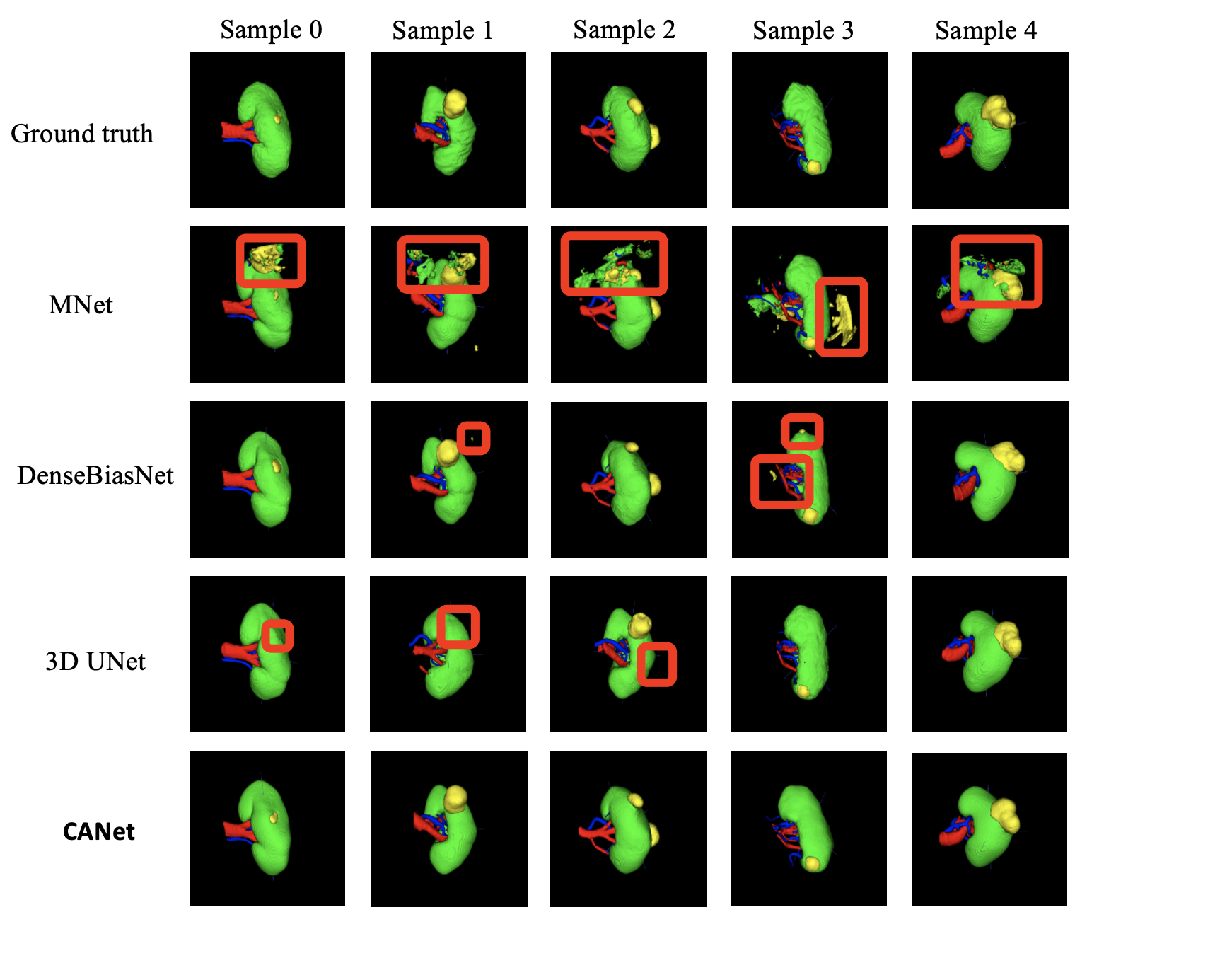}
\label{network.eps}
\caption{The visualization results of the experiments. Five sample were used to visualize the four methods, and the part marked by the red thick frame is the place that is quite different from the ground truth.}
\vspace{-1em}  
\end{center}
\end{figure}

Table 2 demonstrates the qualitative results of our proposed CANet compared to three traditional methods and the nn-UNet backbone. From Figure 2, we can conclude that CANet has the best performance in all results. With regard to DSC, our CANet and nn-UNet backbone have shown excellent performance compared with the other three models. Meanwhile, CANet has improved the results of these four segmentation parts compared with the nn-UNet backbone for 0.4\%, 0.2\%, 0.8\% and 0.8\%. The improvement in veins and arteries is still very obvious. Among them, comparing CANet and nn-UNet backbone, our model has a large improvement in $HD(mm)$, which are respectively improved by 1.63mm, 4.46mm, 8.12mm and 8.17mm.

The findings of the visualization are displayed in Figure 5 below. The other three deep learning approaches do poorly, roughly speaking. Based on the findings of the onion visualization, MNet performs the poorest. DenseBiasNet has the potential to discover more malignancies, but 3D UNet frequently misses tumors. In comparison to existing methods of deep learning, our CANet has obtained superior results in terms of both stereotypes and quantification.

\section{Conclusion}
This paper proposes a simple but effective method named channel extending and axial attention catching Network(CANet) for multi-structure kidney segmentation based on the nn-UNet framework. To improve the performance of the four-item mixed structure, we begin by employing a channel extending method to assist the extraction of complicated structural information. To further improve the edges, we add an axial attention catching module(AAC) to the decoder. After testing on the KiPA22 dataset, the persuasive findings imply that the newly suggested CANet is successful.

%
%

\nocite{*}
\bibliography{myReference}
\bibliographystyle{splncs04}

\end{document}